\documentstyle{article}

\input epsf

\newcommand{\mathbf}{\bf}

\begin{document}

\begin{center}
{\huge\bf Uncertainty Relation For Quantized Magnetic Fields and The Quantum Hall Effect}
\end{center}

\vspace{1cm}
\begin{center}
{\large\bf 
F.GHABOUSSI}\\
\end{center}

\begin{center}
\begin{minipage}{8cm}
Department of Physics, University of Konstanz\\
P.O. Box 5560, D 78434 Konstanz, Germany\\
E-mail: ghabousi@kaluza.physik.uni-konstanz.de
\end{minipage}
\end{center}

\vspace{1cm}

\begin{center}
{\large{\bf Abstract}}
\end{center}

\begin{center}
\begin{minipage}{12cm}
It is shown that the stepped QHE curve which represents the relation between Hall potential $A_H$ and the gate potential $A_G$ can be described by the uncertainty relation $\Delta A_H \cdot \Delta A_G \geq \displaystyle{\frac{\hbar B}{e}}$ which is related with the canonical flux quantization.
\end{minipage}
\end{center}

\newpage
It is known that the classical Coulomb potential is determined by its $\displaystyle{\frac{1}{r}}$ behaviour which is represented by the $\displaystyle{\frac{1}{r}}$ curve. Now, if one investigates the stepped QHE curve \cite{K} which represents the measured Hall potential $A_H$ with respect to the gate potential $A_G$, then one finds out that also this curve represents the {\it quantum} behaviour of the Hall potential with respect to some distance or length. The reason is that the gate potential can be considered here as given by $A_G = B \cdot r$ where $B$ is a constant magnetic field. Thus, here also the QHE curve can be considered as a relation between the {\it quantized} Hall potential and the length $r$.

Recall further that if one planes the same stepped QHE curve, so that all plateau regions become smooth, then this smooth curve takes the shape of a {\it classical} Coulomb potential $A_r = \displaystyle{\frac{1}{r}}$. Thus, in view of the fact that the plateaus manifest the quantum character of Hall potential, then their planing which results in the classical potential curve should be considered as the transition to the classical limit, i. e. to the classical Coulomb potential. In other words, the stepped QHE curve should be considered as if it describes the quantum behaviour of the measured {\it quantized} Hall potential with respect to a finite distance like $\Delta x$ which is related with some measures of plateaus \cite{N}.

For this purpose, i. e. the description of QH behaviour of Hall potential which is a two dimensional quantum effect, we use some recent results from the canonical flux quantization which is also a quantum model for two dimensional electromagnetic field:

Recently, it was shown that the canonical flux quantization $S_{(cl)} ^{(flux)} = \Phi  = \oint e A_m dx^m =$ 

$\int \int e F_{mn} dx^m \wedge dx^n = N h \ , N \in {\mathbf Z} \ ; \ m, n = 1, 2$ requires a new uncertainty relation: $e \Delta A_m \cdot \Delta x_m \geq \hbar$ ( no summation) , where $\Delta A_m$, $\Delta x_m$ and $F_{mn}$ are, respectively, the uncertainty of quantized electromagnetic potential, the correlated position uncertainty and the magnetic field strength \cite{me}.

The reason is that the neccessary comparison between $S_{(cl)} ^{(flux)}$ and the classical canonical action functional $S^{(canon)} _ {(cl)} = \int \int dP_m \wedge dx^m$ to fix the phase space variables of flux system $S_{(cl)} ^{(flux)}$, determines $e A_m$ and  $x_m$ to be the canonical conjugate variables of phase space of flux system \cite{me} \cite{act}. Hence, the commutator postulate for canonical quantization of flux system, should be of the form: $e [ \hat{A}_m  ,  \hat{x}_n ] = -i \hbar \delta_{mn}$. 
Therefore, in view of the well known relation between commutator postulate and uncertainty relation in quantum theory, such a coanonical quantization will require also the existence of an uncertainty relation: $e \Delta A_m  \cdot \Delta x_m \geq \hbar$ \cite{me}.

Recall that the commutator of relative coordinate operators in the cyclotron motion: $[\hat{x}_m \ , \ \hat{x}_n] = -i l_B ^2 \epsilon_{mn} \ , \epsilon_{mn} = - \epsilon_{nm} = 1$ \cite{LA} is proportinal to the commutator $e [ \hat{A}_m  ,  \hat{x}_n ] = -i \hbar \delta_{mn}$ by the Landau gauge $A_m = B x^n \epsilon_{mn}$, where $l_B$ is the magnetic length which is defined by 
$l_B ^2 := \displaystyle{\frac{\hbar}{eB}}$. 
Accordingly, there is an equivalent uncertainty relation $e B \Delta x_m \cdot \Delta x_n \geq \hbar |\epsilon_{mn} |$ in accord with the commutator $[\hat{x}_m \ , \ \hat{x}_n] = -i l_B ^2 \epsilon_{mn}$, that can be obtained from the original uncertainty relation by the "Landau gauge" $\Delta A_m = B \cdot \Delta x^n | \epsilon_{mn} |$.

Recall also that the commutator in the cyclotron motion can be rewritten as $ e B [\hat{x}_m \ , \ \hat{x}_n] = -i \hbar \epsilon_{mn}$ which is the quantization commutator postulate that manifests the canonical flux quantization: 

$\int \int e F_{mn} dx^m \wedge dx^n = N h$, whereas the commutator $e [ \hat{A}_m  ,  \hat{x}_n ] = -i \hbar \delta_{mn}$ manifests the equivalent canonical flux quantization: $\oint e A_m dx^m = N h$ \cite{zwischen}. Therefore, in view of the existence of related uncertainty relations $e \Delta A_m . \Delta x_m \geq \hbar$ and $e B \Delta x_m . \Delta x_n \geq \hbar |\epsilon_{mn} |$, the above introduced "Landau gauge" relation $\Delta A_m \geq B \cdot \Delta x^n | \epsilon_{mn} |$ can be considered as a quantum consistency relation between them, which should be valid in relevant quantum cases.

We will show that the mentioned QHE curve is described, at least in its schematic stepped form \cite{qhe}, by the "uncertainty equations" $\Delta A_m = \displaystyle{\frac{\hbar}{e \Delta x_m}}$ or $e B \Delta x_m \cdot \Delta x_n = \hbar |\epsilon_{mn}|$ with $( \Delta x_m )_{(minimum)} = l_B$, which are special cases of the discussed uncertainty relations.
Thus, the actual stepped QHE curve \cite{K} should be described by the uncertainty relation 
$\Delta A_m \geq \displaystyle{\frac{\hbar}{e \Delta x_m}}$.

\bigskip
First let us mention that the introduced uncertainty equation $e \Delta A_m \cdot l_B = e B \cdot l_B ^2 = \hbar$ is itself a consistent relation, in view of the fact that it matches with the independent definition of magnetic length: $l_B ^2 := \displaystyle{\frac{\hbar}{eB}}$. 

Moreover, since the equation $A_m = \displaystyle{\frac{1}{x_m}}$ describes a curve, then the equations $\Delta A_m = \displaystyle{\frac{1}{\Delta x_m}}$ or $\Delta A_m = \displaystyle{\frac{\hbar}{e \Delta x_m}}$ should describe also a "curve" which has however a finite character in accord with the finiteness of $\Delta A_m$ and $\Delta x_m$.  

\bigskip
As a consistency argument that such an uncertainty relation should be the right relation to describe the QHE, let us mention also that the uncertainty equation $e B \Delta x_m . \Delta x_n = \hbar |\epsilon_{mn}|$ for $\Delta x_m = l_B$, i. e. $e B \cdot l^2 _B = \hbar$ can be rewritten as $\rho_H = \displaystyle{\frac{1}{\nu}\frac{h}{e^2}}$ which describes the schematic QHE curve $\rho_H$ with respect to $\nu$. 
Whereby $\nu = 2 \pi n l_B ^2 \ , \ \rho_H = \displaystyle{\frac{B}{ne}}$ and $n$ are, respectively, the filling factor, the Hall resistivity and the global electronic density on the QHE sample.

\newpage
In other words, we will show that in the same manner that the curve $ \displaystyle{\frac{1}{x}}$ represents the {\it classical} Coulomb potential $A_x = \displaystyle{\frac{e}{x}}$ for $e := 1$, so the schematic stepped QHE curve represents the {\it quantum} potential $\Delta A_H = \displaystyle{\frac{\hbar}{e \Delta x_H}}$, where $\Delta A_H$ and $\Delta x_H$ are the quantum uncertainty for the Hall potential and the position uncertainty in the Hall direction on the sample. Thereby, we show that the uncertainty relation $ e \Delta A_H \cdot \Delta x_H \geq \hbar$ is equivalent to the uncertainty relation $(\Delta A_H) \cdot (\Delta A_G) \geq \displaystyle{\frac{\hbar B}{e}}$ which should describe also the QH relation between  Hall- and the gate potential.

Recall also that, in the classical (Coulomb) case $A_x = \displaystyle{\frac{e}{x}}$ is a function of $x$, whereas in the quantum case $\Delta A_x = \displaystyle{\frac{\hbar}{e^2}} \displaystyle{\frac{e}{\Delta x}}$ there is only a dependency between the finite values $\Delta A_x$ and $\Delta x$. 

Furthermore, from the simple algebra we know that the coordinates of any point $p$ on the curve $A_x = \displaystyle{\frac{1}{x}}$ fulfil the relation $(A_x)_p \cdot (x)_p = 1$. Hence, the coordinates of any {\it corner} point $P$ on the quantum "curve" $\Delta A_x = \displaystyle{\frac{\hbar}{e \Delta x}}$ should fulfil just the relation $(\Delta A_x)_P \cdot (\Delta x)_P = \displaystyle{\frac{\hbar}{e}}$, where $\Delta A_x$ and $\Delta x$ are the finite variation of point $P$ in both directions on this curve parallel to the $A_x$- and  $x$ axis (see fig. 1) \cite{erk}. 

Moreover, the {\it schematic} areas of cells between the curves $\Delta A_x = \displaystyle{\frac{\hbar}{e \Delta x}}$ and $A_x = \displaystyle{\frac{1}{x}}$
are about the half of product $\Delta A_x \cdot \Delta x$, i. e. $\displaystyle{\frac{\hbar}{e}}$ (see fig. 1) \cite{erk}.
Therefore, in the classical limit, where $\Delta A_x \cdot \Delta x = \displaystyle{\frac{\hbar}{e}} \rightarrow 0$ the curve $\Delta A_x = \displaystyle{\frac{\hbar}{e \Delta x}}$ approaches to the curve $A_x = \displaystyle{\frac{1}{x}}$. We will show that the same relations is fulfilled, at least, for the schematic QHE curve. 

To see the relation between the curve $\Delta A_x = \displaystyle{\frac{\hbar}{e \Delta x}}$ and the schematic QHE curve \cite{qhe} recall that on the one hand, the schematic QHE curve of Hall potential $A_H$ with respect to the gate potential $A_G$ describes also the quantized values $\rho_H = \displaystyle{\frac{h}{e^2 \nu}}$. On the other hand, the same equation $\rho_H = \displaystyle{\frac{1}{\nu}} \displaystyle{\frac{h}{e^2}}$ can be obtained from the equation $B = \displaystyle{\frac{\hbar}{e l_B ^2}}$ and this one can be obtained from the uncertainty equation $\Delta A_x = \displaystyle{\frac{\hbar}{e \Delta x}}$ where $\Delta x = l_B$, by $\Delta A_x = B \cdot l_B$.

Therefore, the uncertainty equation $\Delta A_x = \displaystyle{\frac{\hbar}{e \Delta x}}$ should describe the schematic QHE curve, if we identify the $x$- and the $y$-direction with the Hall- and the gate direction, respectively. Thus, according to $\Delta A_G = B \cdot \Delta x = B . \Delta x_H$ and $\Delta A_H = \Delta A_G = B \cdot l_B$, one obtains from the uncertainty equation $\Delta A_x = \displaystyle{\frac{\hbar}{e \Delta x}}$ the equation $\Delta A_H = \displaystyle{\frac{\hbar B}{e \Delta A_G}}$ which should describe also the schematic QHE curve. 

Hereby, the potential uncertainties $\Delta A_H$ and $\Delta A_G$ should be considered as the heigth and length of steps on the QHE curve, i. e. they should represent the height between two subsequent plateaus and the width of plateaus. 

\newpage
Hence, for every corner point $P$ on the schematic QHE curve, following relations will hold: 

$(\Delta A_H)_P \cdot (\Delta A_G)_P = \displaystyle{\frac{\hbar B}{e}}$, $(\Delta A_H)_P . (\Delta x_H)_P = \displaystyle{\frac{\hbar}{e}}$ or $(B)_P . (l_B ^2)_P = \displaystyle{\frac{\hbar}{e}}$ \cite{furt}. 

In other words, these equations between different but related quantities describe the same schematic QHE curve. This is in accord with the fact that the QHE can be represented by similar curves, e. g. as the relation between $(\rho_H \ , \ B) \ , \ ( A_H \ , \ A_G)$ or $( \sigma_H \ , \ \nu)$, etc. \cite{qhe}.

Moreover, although an exact agreement will exists only between the uncertainty equation and the schematic QHE curve for an ideal QHE \cite{2K}, where for example the electronic current flows in a $\Delta x = l_B$ distance  
from the edge of sample. However, as it is mentioned above, one should  obtain a good agreement between the uncertainty relation $\Delta A_H \cdot \Delta A_G \geq \displaystyle{\frac{\hbar B}{e}}$ and the actual QHE curve \cite{K}. Hence also the curve $\Delta A_H = \displaystyle{\frac{\hbar B}{e \Delta A_G}}$, which is the same curve as $\Delta A_H = \displaystyle{\frac{\hbar}{e \Delta x_H}}$, approximates the ("classical") $A_H = \displaystyle{\frac{1}{x_H}}$ curve (see fig. 1).

For the case $(\Delta A_H \cdot \Delta A_G) \rightarrow 0$, which is achived for $\hbar \rightarrow 0$ or even for $B \rightarrow 0$ (the classical state), the stepped QHE curve approches the continuous $A_H = \displaystyle{\frac{1}{x_H}}$ curve.  

\bigskip
The conclussion is that the canonical flux quantization $\Phi = e \oint A_H dx_H = N h$ on the QH sample, which is described also by the commutator postulate $e [ \hat{A}_H  ,  \hat{x}_H ] = -i \hbar$ or by the uncertainty relation $ e \Delta A_H \cdot \Delta x_H \geq \hbar$, should describe also the QHE as the quantum behaviour of Hall potential with respect to the gate potential in the following way:

In view of the fact that for {\it any} point $P$ on the actual stepped curve \cite{K} one of the correlated uncertainty relations: $(\Delta A_H)_P \cdot (\Delta A_G)_P \geq \displaystyle{\frac{\hbar B}{e}}$ or $(\Delta A_H)_P \geq B \cdot l_B$ or $(\Delta A_G)_P \geq B \cdot l_B$ should be fulfilled, therfore the mentioned QHE curve should be described by the following uncertainty relations:

\begin{eqnarray}
&& \Delta A_H \cdot \Delta A_G \geq \displaystyle{\frac{\hbar B}{e}}\nonumber\\ 
&& \Delta A_H \geq B \cdot l_B\nonumber\\ 
&&\Delta A_G \geq B \cdot l_B 
\end{eqnarray}

\newpage
Thus, in view of $\Delta A_G = B \cdot \Delta x_H$ the same curve should be described also by the uncertainty relations:

\begin{eqnarray}
&& \Delta A_H \cdot \Delta x_H \geq \displaystyle{\frac{\hbar}{e}}\nonumber\\ 
&& \Delta A_H \geq B \cdot l_B\nonumber\\ 
&& \Delta x_H \geq l_B 
\end{eqnarray}

\bigskip

\bigskip 

\bigskip

\bigskip

\bigskip

\bigskip

\bigskip

\bigskip

\bigskip

\bigskip

\bigskip

\bigskip

\bigskip

\bigskip

\bigskip

\bigskip

\bigskip

\bigskip

\bigskip

\bigskip

\unitlength 1cm
\begin{flushleft}

\begin{picture}(15,7.5)




\put(3.2,1){\epsfxsize=10cm \epsfbox{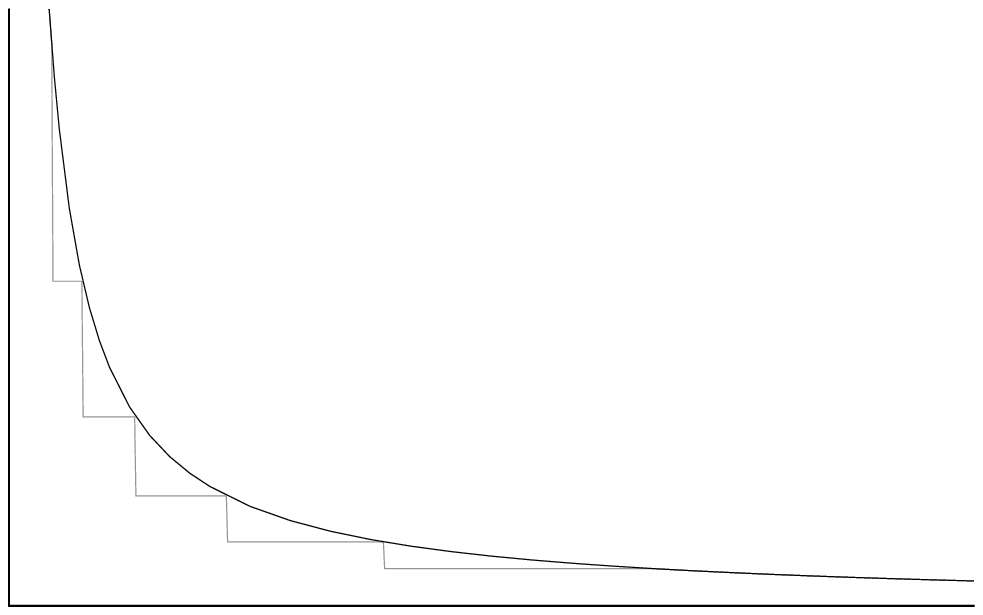}}
\put(4.3,2.18){\epsfysize=0.9cm \epsfbox{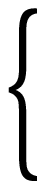}}
\put(4.53,1.97){\epsfxsize=1.05cm \epsfbox{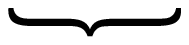}}
\put(3.2,7.3){$\displaystyle A_x$}
\put(13,0.9){x}
\put(6,6.5){$  \displaystyle \left( A_x= \frac{1}{x}\right) \quad \hbox{or} \quad  \left( A_H=\frac{1}{x\scriptstyle{_H}}\right) $}
\put(5.9,6.5){\vector(-3,-2){2}}
\put(4.9,1.8){$\displaystyle \Delta$x}
\put(3.55,2.6){$\displaystyle \Delta A_x$}
\put(0.5,5.3){$\displaystyle \left( \Delta A_x = \frac{1}{ \Delta x}\right)$}
\put(0.1,4.3){$\displaystyle \hbox{or} \left( \Delta A_H = \frac{\hbar}{e\, \Delta x\scriptstyle{_H}}\right) $}
\put(2.9,5.1){\vector(3,-1){0.8}}
\end{picture}
\end{flushleft}

Fig. 1: The schematic curves $\Big((A_x = \displaystyle{\frac{1}{x}})$ or ($A_H = \displaystyle{\frac{1}{x_H}}) \Big)$ and $\Big((\Delta A_x = \displaystyle{\frac{1}{\Delta x}})$ or ($ \Delta A_H = \displaystyle{\frac{\hbar}{e \Delta x_H}}) \Big)$

\newpage
Footnotes and references


\begin{thebibliography}{100}
\bibitem{N}
It can describe also the quantum behaviour of some related quantity to Hall potential, e. g. the Hall resistivity, with respect to a correlated quantity to the mentioned distance, e. g. the applied magnetic field $B$.
\bibitem{erk}
These curves are shematic curves and the units are renormalized to make a comparison between the continuous and stepped curves possible.
\bibitem{furt}
Further recall that the experimental variation of values of $B$ or of density $n$ in QHE experiments varies also the value of $l_B$.
\end{thebibliography}
\end{document}